\documentclass[twocolumn,aps,pra,showpacs,raggedbottom,nobalancelastpage,amssymb,superscriptaddress,longbibliography]{revtex4-1}

\usepackage{graphicx}
\usepackage{amsmath}
\usepackage{amssymb}
\usepackage{bm}
\usepackage[usenames]{color}
\usepackage{amsfonts}

\usepackage{comment}

\usepackage[colorlinks=true , citecolor=blue,urlcolor=blue]{hyperref}

\renewcommand{\i}{\textrm{i}}

\def\kb{{\mathchar'26\mkern-9mu k}}

\usepackage{ulem}

\include{./journalsVaps}

\begin{document}
\title{Dynamical localization of interacting bosons in the few-body limit}
\author{Radu Chicireanu}
\affiliation{Universit\'e  de  Lille,  CNRS,  UMR  8523  --  PhLAM  --  Laboratoire  de Physique  des  Lasers,  Atomes  et  Mol\'ecules,  F-59000  Lille,  France}
\author{Adam Ran\c con}
\affiliation{Universit\'e  de  Lille,  CNRS,  UMR  8523  --  PhLAM  --  Laboratoire  de Physique  des  Lasers,  Atomes  et  Mol\'ecules,  F-59000  Lille,  France}

\begin{abstract}
The quantum kicked rotor is well-known to display dynamical localization in the non-interacting limit. In the interacting case, while the mean-field (Gross-Pitaevskii) approximation displays a destruction of dynamical localization, its fate remains debated beyond mean-field. Here we study the kicked Lieb-Liniger model in the few-body limit. We show that for any interaction strength, two kicked interacting bosons always dynamically localize, in the sense that the energy of the system saturates at long time. However, contrary to the non-interacting limit, the momentum distribution $\Pi(k)$ of the bosons is not exponentially localized, but decays as $\mathcal C/k^4$, as expected for interacting quantum particles, with Tan's contact $\mathcal C$ which remains finite at long time. We discuss how our results will impact the experimental study of kicked interacting bosons.  
\end{abstract}

\date{\today}
\maketitle

\section{Introduction}

The Quantum Kicked Rotor (QKR) is a paradigmatic model of  quantum chaos. It is most famous for displaying dynamical localization, which is the analog of Anderson localization in momentum space \cite{Fishman1982}. Experimental realizations of the atomic QKR and its variants have allowed for detailed studies of Anderson localization and two dimensions \cite{Manai2015}, the Anderson transition in three dimensions \cite{Chabe2008}, as well as the study of the effects symmetries on weak localization  \cite{Hainaut2018b} and classical-to-quantum transition at early times \cite{Hainaut2018a}.

The effects of inter-atomic interactions on dynamical localization is an intriguing problem. Indeed, because localization is in momentum space but interactions are effectively local in real space (hence, long-range in momentum), the interacting QKR is expected to behave differently from a standard disordered interacting quantum system. In the latter case, strong enough disorder is known to produce, at least in low dimensions, a new phase of matter, the many-body localized (MBL) phase \cite{Nandkishore2015,Abanin2019}. This phase is not ergodic and does not allow for thermalization. In particular, driven MBL system can resist heating, in contrast with the expectation of heating to infinite temperature for delocalized phases of interacting systems \cite{Ponte2015a} (however, for a counter-example, see e.g. \cite{Chandran2016}).

This therefore raises the question of the existence of a many-body dynamically localized (MBDL) phase in the interacting QKR. There have been studies for various toy-models \cite{Adachi1988,Borgonovi1995,WenLei2009,Keser2016,Rozenbaum2017}, as well as for more realistic models for cold atoms. At the mean-field level, it has been argued both on theoretical and numerical grounds that the interactions will destroy dynamical localization, which is replaced by a subdiffusion in momentum space \cite{Shepelyansky1993,Pikovsky2008,Flach2009,Gligoric2011,Cherroret2014,Lellouch2020}. Recently, the study of more realistic models of interacting atomic bosons periodically kicked, the kicked Lieb-Liniger model, have led to seemingly contradictory results. Using various many-body techniques, Rylands et al. \cite{Rylands2020} have argued that the system should not heat up, thus leading to MBDL. On the other hand, Qin et al. \cite{Qin2017} have studied the kicked Lieb-Liniger model with only two particles, which allows for a more exact treatment of the problem. There, they have found that the energy of the system seems to increase, indicating a breakdown of dynamical localization.

In this paper, we revisit the dynamics of two interacting bosons described by the kicked Lieb-Liniger model. We analyze in details the dynamics of the system and show that the energy always saturates at long times for any interaction strength. This indicates that the system is indeed localized dynamically. However, we show that the momentum distribution of the system, which is a quantity directly accessible in ultracold atomic gas experiments, \textit{does not} decay exponentially at large momenta $k$ as for non-interacting particles, but as a power law $k^{-4}$ as expected for interacting quantum systems. The manuscript is organized as follows: we introduce the model in Sec.~\ref{sec_model} and discuss the dynamics in Sec.~\ref{sec_dyn}. We analyze the momentum distribution and give a quantitative description in the infinite interaction limit in Sec.~\ref{sec_Pik}. Finally, we discuss our results in Sec.~\ref{sec_discuss}.

\section{The interacting quantum kicked rotor \label{sec_model}}

We study two interacting bosons in a ring of circumference $L=2\pi$, with Hamiltonian $\hat H = \hat H_{LL}+ \hat H_{K}$. Here $\hat H_{LL}$ describes the dynamics of the interacting bosons between the kicks, and is given by the Lieb-Liniger Hamiltonian~\cite{Lieb1963}
\begin{equation}
\hat H_{LL} = \frac{\hat p_1^2}{2}+\frac{\hat p_2^2}{2}+g\, \delta(\hat x_1-\hat x_2),
\end{equation}
and the kick Hamiltonian reads
\begin{equation}
\hat H_{K}=K\left(\cos(\hat x_1)+\cos(\hat x_2)\right)\sum_n \delta(t-n).
\end{equation}
We use the standard units of the (non-interacting) kicked rotor: time is in units of the kick period $T$, positions are in units of $L/2\pi$  (which is also the inverse wavevector of the kicking potential), and momenta are in units of $M L/T$, with $M$ the mass of the bosons. The canonical commutation relations are then given by $[\hat x_i, \hat p_j]=\delta_{ij}\kb$, with $\kb=\frac{4\pi^2 \hbar T}{M L^2}$ the effective Plank constant~\cite{Lemarie2009}. The dimensionless interaction strength $g$ is related to the one-dimensional scattering length $a$ by $g=-\frac{L}{a}\frac{\kb^2}{4\pi^3}$ \cite{Olshanii2003}.

To study the dynamics of the system, it is convenient to use the eigenbasis of the Lieb-Liniger Hamiltonian. Following Lieb and Liniger, it is easily found using a Bethe ansatz, and the eigenfunctions of $\hat H_{LL}$ read 
\begin{equation}
\Phi^n_{m}(x_1,x_2)=\frac{ e^{i \frac n2 (x_1+x_2)}}{\sqrt{2\pi}}\frac{\sin\left(k_m |x_1-x_2|-\frac{\theta_m}2\right)}{\sqrt{\pi-\frac{\sin(\theta_m)}{2k_m}}}.
\end{equation}
Here, $n\in\mathbb{Z}$ is the momentum of the center-of-mass (in units of $\kb$). The relative momentum $k_m=\frac{m+\theta_m/\pi}{2}$ (in units of $\kb$) is parametrized by a positive integer $m$, and the phase-shift induced by the interaction $\theta_m$. The periodic boundary conditions and the delta-interaction give the constraints that $m+n$ must be odd, and 
\begin{equation}
 \theta_m = -2 \arctan\left(\frac{2 \kb^2 k_m}{g}\right).
\end{equation} 
The energy of the state $|\Phi^n_{m}\rangle$ is $E^n_m=\frac{\kb^2}{4}(n^2+4 k_m^2)$.

\begin{figure}[t!]
	\centering
	\includegraphics[width=\columnwidth]{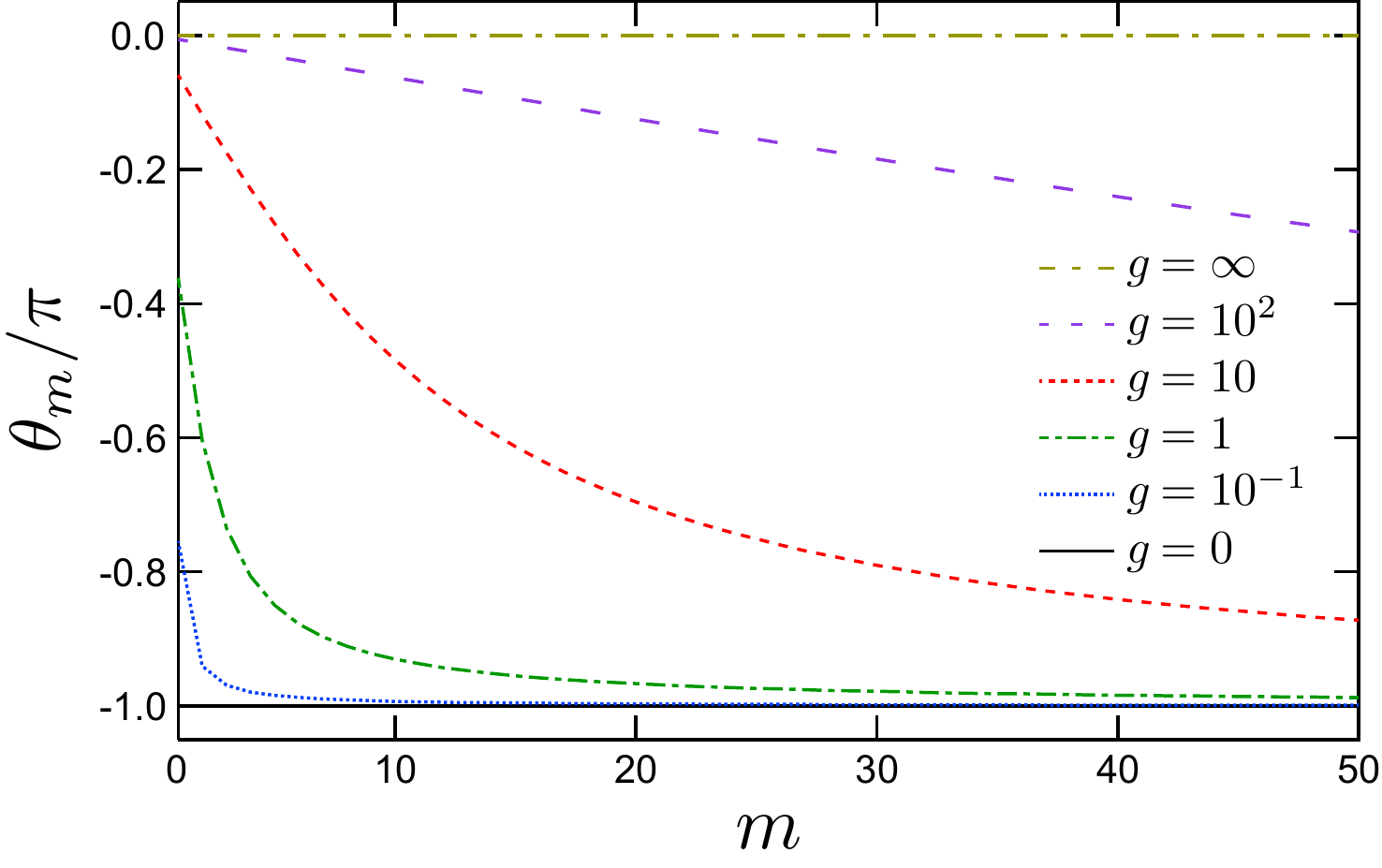}
	\caption{Phase-shift $\theta_m$ as a function of $m$ for different values of the interaction parameter $g$ ($\kb=1$). The value of $g$ increases from bottom to
		top curve.}
	\label{fig1_theta} 
\end{figure}

The phase-shift $\theta_m$ is shown in Fig.~\ref{fig1_theta} for different values of the interaction strength, and $\kb=1$. It interpolates between $0$ for small $m$, where the wave function effectively fermionizes, and $\theta_m\to -\pi$ as $m\to \infty$, where the bosons are almost free, as the (relative) kinetic energy dominates over the interaction. In the Tonks limit, $g\to\infty$, $\theta_m=0$ and we recover the Tonks-Girardeau (TG) wave functions \cite{Tonks1936,Girardeau1960}.

The evolution operator over one period is given by
\begin{equation}
\hat U = e^{-i\frac{\hat H_K}\kb}e^{-i\frac{\hat H_{LL}}\kb},
\end{equation}
and its matrix elements read
\begin{equation}
U_{mp}^{nq}\equiv \langle \Phi^n_m|\hat U|\Phi^q_p\rangle = e^{-i \frac{E^n_m}\kb}  \langle \Phi^n_m|e^{-i\frac{\hat H_K}\kb}|\Phi^q_p\rangle.
\end{equation}
The matrix elements of kick operator must be computed numerically for finite $g$, and are given explicitly by
\begin{equation}
  \langle \Phi^n_m|e^{-i\frac{\hat H_K}\kb}|\Phi^q_p\rangle = \int_0^{2\pi}F_{q-n}(x)\psi_p(x)\psi_m(x),
 \end{equation} 
with $\psi_m(x)=\frac{\sin\left(k_m x-\frac{\theta_m}2\right)}{\sqrt{\pi-\frac{\sin(\theta_m)}{2k_m}}}$ and $F_{n}(x)=(-i)^{n}J_{n}\left(\frac{2K}{\kb}\cos\left(\frac x2\right)\right)$, where $J_\nu(z)$ is the $\nu$-th Bessel function of the first kind. 
The asymptotic behavior of these matrix elements has been analyzed in Ref.~\cite{Qin2017}. There, it has been shown that for fixed $m$ and $p$, $|U_{mp}^{nq}|$ decays as $(|n-q|!)^{-1}$, much faster than an exponential, while at fixed $n,q,p$, it decays as $m^{-4}$~\footnote{This result can be generezalized to show that at fixed $n,q$, the matrix elements decay as $(m^2-p^2)^2$ for sufficiently large $m$ and $p$.}. This power law decay has been interpreted by the authors of Ref.~\cite{Qin2017} to be the cause of the breakdown of dynamical localization in this model, see however the discussion of this argument in Sec.~\ref{sec_discuss}.

To compute the time evolution of the system, we expand its wave function in the Lieb-Liniger basis, $|\Psi_t\rangle = \sum_{n,m}c^n_m(t) |\Phi^n_m\rangle$, where the coefficients $c^n_m(t)$ obey the stroboscopic evolution $c^n_m(t+1)=\sum_{q,p}U_{mp}^{nq}c^q_p(t)$. Here and in the following, we always assume that the sum is performed over the allowed values of $m$ and $n$ ($m\in \mathbb{N}^*$, $n\in \mathbb Z$ and $n+m$ odd).  To perform the time-evolution numerically, it is necessary to truncate the basis, and we only keep states with $|n|\leq n_{max}$ and $m\leq m_{max}$, with typical values of $n_{max}=160$ and $m_{max}=160$. We have checked that these values used in our numerics are such that our results are converged, in the sense that physical observables do not change when $n_{max}$ and $m_{max}$ are increased, and the that the normalization of the wave function stays very close to one at all times (such that the states $|\Phi^n_m\rangle$ with $n>n_{max}$ and $m>m_{max}$ would not be significantly populated if they were included). Here and in the following, we will always assume that the system starts in the groundstate of the Lieb-Liniger Hamiltonian, $|\Psi_{t=0}\rangle = |\Phi_1^0\rangle$. We use $K=3$ and $\kb=1$ in the numerics, which allows us to use a not too large basis. 

One difficulty in the study of the dynamics of this problem is that the various observables typically display large fluctuations during time-evolution. This also happens in the context of the QKR, and in that case, one usually averages over the quasi-momentum $\beta$, which is a dynamically conserved quantity. Changing the quasi-momentum there corresponds to a change of the disorder realization of the corresponding Anderson model~\cite{Lemarie2009}. In order to simplify the analysis of our numerics, we introduce an artificial ``quasi-momentum'' in the energy of the Lieb-Liniger model, i.e. we replace $E^n_m$ by $E^{n+2\beta}_m$, equivalent to add a magnetic flux in the system. This way of introducing the quasi-momentum is consistent with what is done in the non-interacting limit. In practice, we average typically over 100 and 500 values of $\beta$ sampled uniformly in $[0,1/2]$, and write the average of an observable $O$ by an overline, $\overline{O}$.

\section{Dynamical localization of interacting bosons \label{sec_dyn}}

\begin{figure}[t!]
	\centering
	\includegraphics[width=\columnwidth]{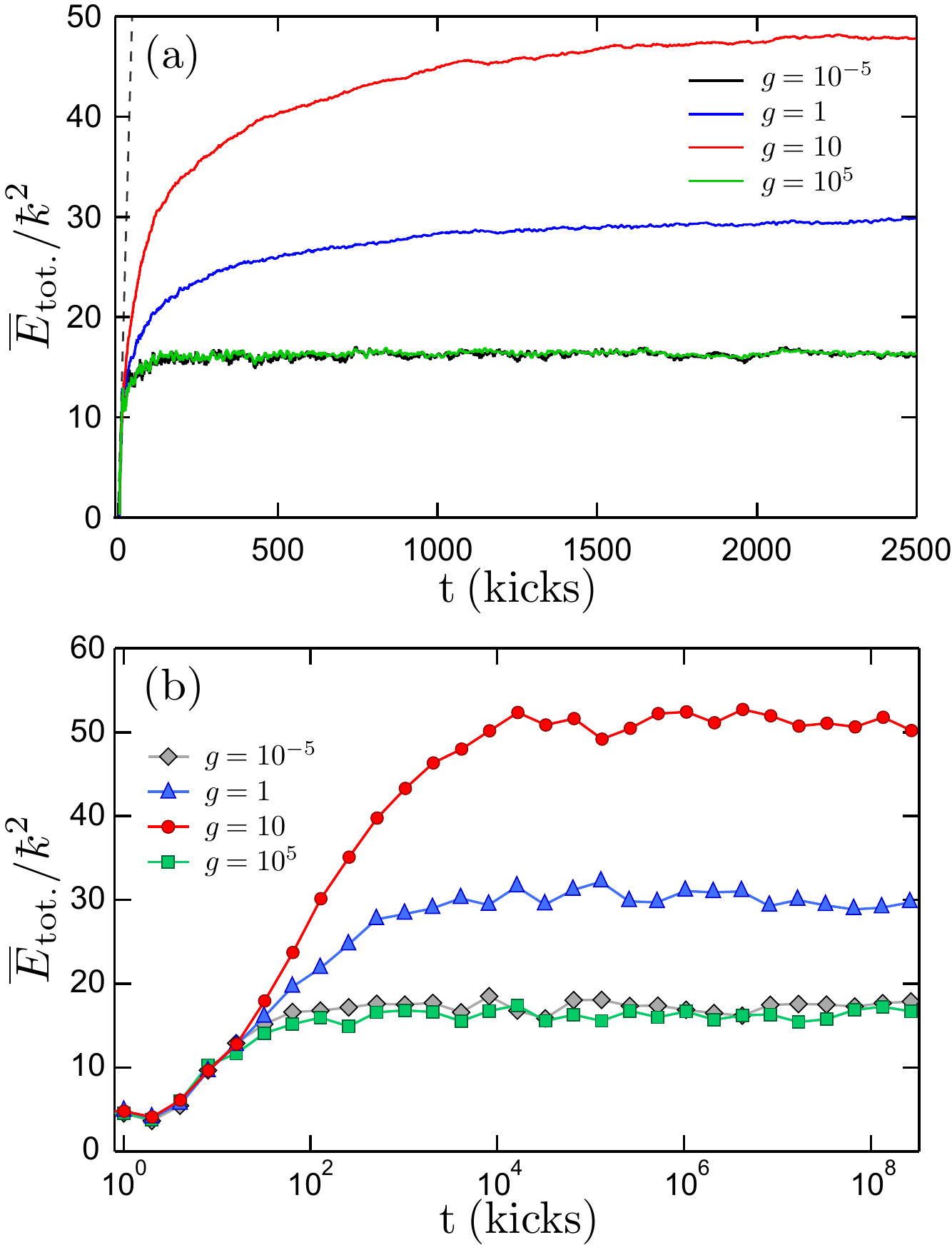}
	\caption{Evolution of the averaged energy of the system, showing a saturation at long times for different values of the interaction strength $g$ for $K=3$, $\kb=1$, in linear (a) and semi-logarithmic (b) scale. Curves in (a) correspond, from top to bottom, to $g=10$, $1$, $10^{-5}$ and $10^{5}$ respectively.}
	\label{fig2_ev_E} 
\end{figure}

The top panel of Fig.~\ref{fig2_ev_E} shows the time-evolution of the energy of the system $\overline E_{\rm tot.}(t)=\overline{\langle\Psi_t|\hat H_{LL}|\Psi_t\rangle}$ for various values of $g$, up to 2500 kicks. We observe a behavior similar to that of the dynamical localization of the non-interacting QKR: at very short times, the energy increases linearly, with a rate independent of $g$ (dashed line) -- which hints that the classical diffusion constant might be rather insensitive to interactions. This initial behavior is followed by a decrease of diffusion and ultimately by a saturation of the energy. We conclude that, even in presence finite interactions, the system does not heat to infinite energy, which is a hallmark of localization for interacting system. In this sense, the system dynamically localizes. 

To check that the system \textit{does} truly localize asymptotically (i.e., that delocalization of the energy does not happen at longer time scales), we have computed the energy after $2^N$ kicks, with $N$ up to 28, by computing $(\hat U)^{2N}$. The bottom panel of Fig.~\ref{fig2_ev_E}  shows that the total energy of the system indeed saturates to a finite value and no sub-diffusive behavior seems to occur even at very large kick numbers. For some finite values of $g$, the localization time (i.e. the time needed for the full saturation of the energy) is significantly longer than in the non-interacting case.  Finally, we have checked that the wave-function coefficients $c_m^n(t)$ do converge at long times to a finite steady-state value.

\begin{figure}[t!]
	\centering
	\includegraphics[width=\columnwidth]{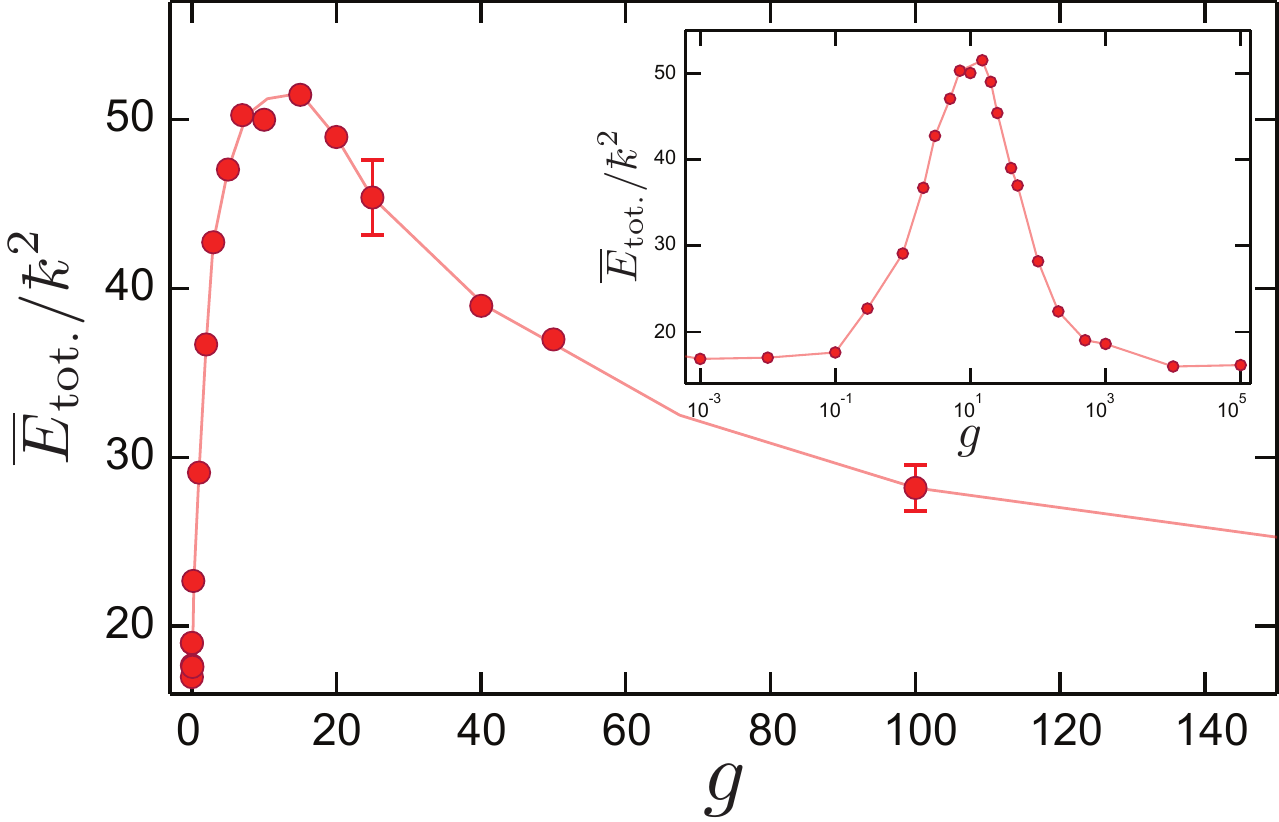}
	\caption{Energy in the localized regime ($t\simeq10^4$ kicks) as a function of $g$ for $K=3$, $\kb=1$. It displays a clear maximum around $g\approx 10$, and decreases towards roughly the same asymptotic values, in the non-interacting (free-bosons, $g\to 0$) and strong-interaction (Tonks-Girardeau, $g\to \infty$) limits. The inset represents the same data, in semi-logarithmic scale for $g$. Statistical error bars are due to averaging over $\beta$ (typically 100 values). The solid line is a guide to the eye.}  
	\label{Fig3-Etot_vs_c} 
\end{figure}

\begin{figure}[t!]
	\centering
	\includegraphics[width=\columnwidth]{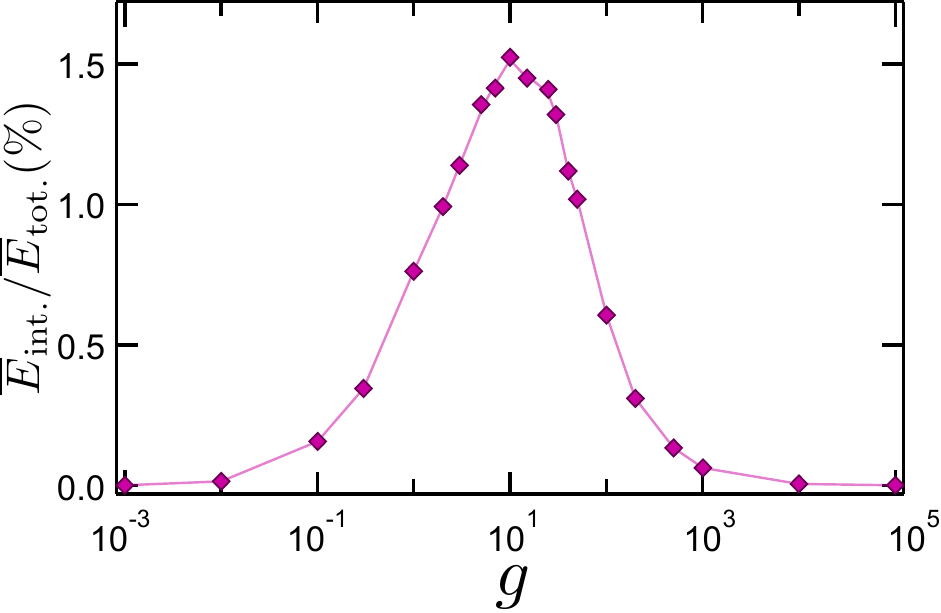}
	\caption{Ratio between the interaction energy and the total energy in the localized regime as a function of the interaction strength $g$ for $K=3$, $\kb=1$. The ratio is at most of $1.5\%$, meaning that most of the energy is in the kinetic energy. The line is a guide to the eye.}
	\label{fig_Eratio} 
\end{figure}

We now proceed to analyze the dependence of various observables as a function of the interaction strength $g$. Fig.~\ref{Fig3-Etot_vs_c} shows the total energy at long-time $  \overline E_{\rm tot.} =\lim_{t\to\infty}  \overline E_{\rm tot.}(t)$ as a function of $g$.  We observe a non-monotonous dependence of the energy as a function of the interaction. This is not too surprising, since in both limits $g=0$ and $g=\infty$, the energy is given by that of non-interacting quasi-particles. In the non-interacting limit, the two bosons start in the zero-momentum state and localize with the same wave function described by the non-interacting QKR. In the opposite limit $g=\infty$, the Tonks limit, the system can be described in terms of non-interacting fermions \cite{Tonks1936,Girardeau1960}. In particular, the energy of the Tonks gas is given by the kinetic energy of those free fermions. The fermions start in the state $\pm\frac{1}{2}$ and localize with wave functions described by the {\it same} localization length $p_{\rm loc}$ (and hence the same final kinetic energy) as the free bosons. Moreover, because the interaction energy also vanishes in the Tonks limit due to the fermionization of the bosons, we therefore expect this two limits to have roughly the same total energy in the long-time limit. Fig.~\ref{fig_Eratio} shows the ratio between the interaction energy, $\overline{E}_{\rm pot.}=\lim_{t\to \infty}\overline{\langle\Psi_t|g\,\delta(\hat x_1-\hat x_2)|\Psi_t\rangle}$, and the total energy in the localized regime. The interaction energy corresponds to a very small contribution, at most $1.5\%$ for $g\simeq 10$, to the total energy, which is therefore dominated by the kinetic energy. The interaction energy vanishes both in the non-interacting limit $g\to 0$ and in the Tonks regime $g\to\infty$, due to the fermionization of the bosons.

\section{Momentum distribution of the dynamically localized Lieb-Liniger gas \label{sec_Pik}}

We shall now address  the momentum distribution $\Pi_t(k)$ of the interacting system, which is a relevant quantity for experiments, and point out key differences with respect to the non-interacting case. The momentum distribution $\Pi_t(k)$ of the system is the Fourier transform of the one-body reduced density matrix (OBRDM) $\rho_t(x,y)$,
\begin{equation}
\Pi_t(k) = \frac{1}{2\pi}\int_0^{2\pi} dx\, \int_0^{2\pi} dy\, e^{i k (x-y)} \rho_t(x,y),
\end{equation}
with the momentum (in unit of $\kb$) $k\in \mathbb{Z}$ due to the periodic boundary conditions, and where the OBRDM is defined as:
\begin{equation}
\rho_t(x,y)=2\int_0^{2\pi}dz \,\Psi^*_t(x,z)\Psi_t(y,z).
\end{equation}
It is normalized such that $\int_0^{2\pi} dx \rho(x,x) = 2$ is the number of particles of the system.  For a given state $|\Psi_t \rangle$, the momentum distribution $ \Pi_t(k)$ is such that $\sum_k \Pi_t(k)=2$ and $\sum_k \frac{\kb^2 k^2}{2} \Pi_t(k)=E_{\rm kin.}(t)=E_{\rm tot.}(t)-E_{\rm int.}(t)$.

\begin{figure}[t!]
	\centering
	\includegraphics[width=\columnwidth]{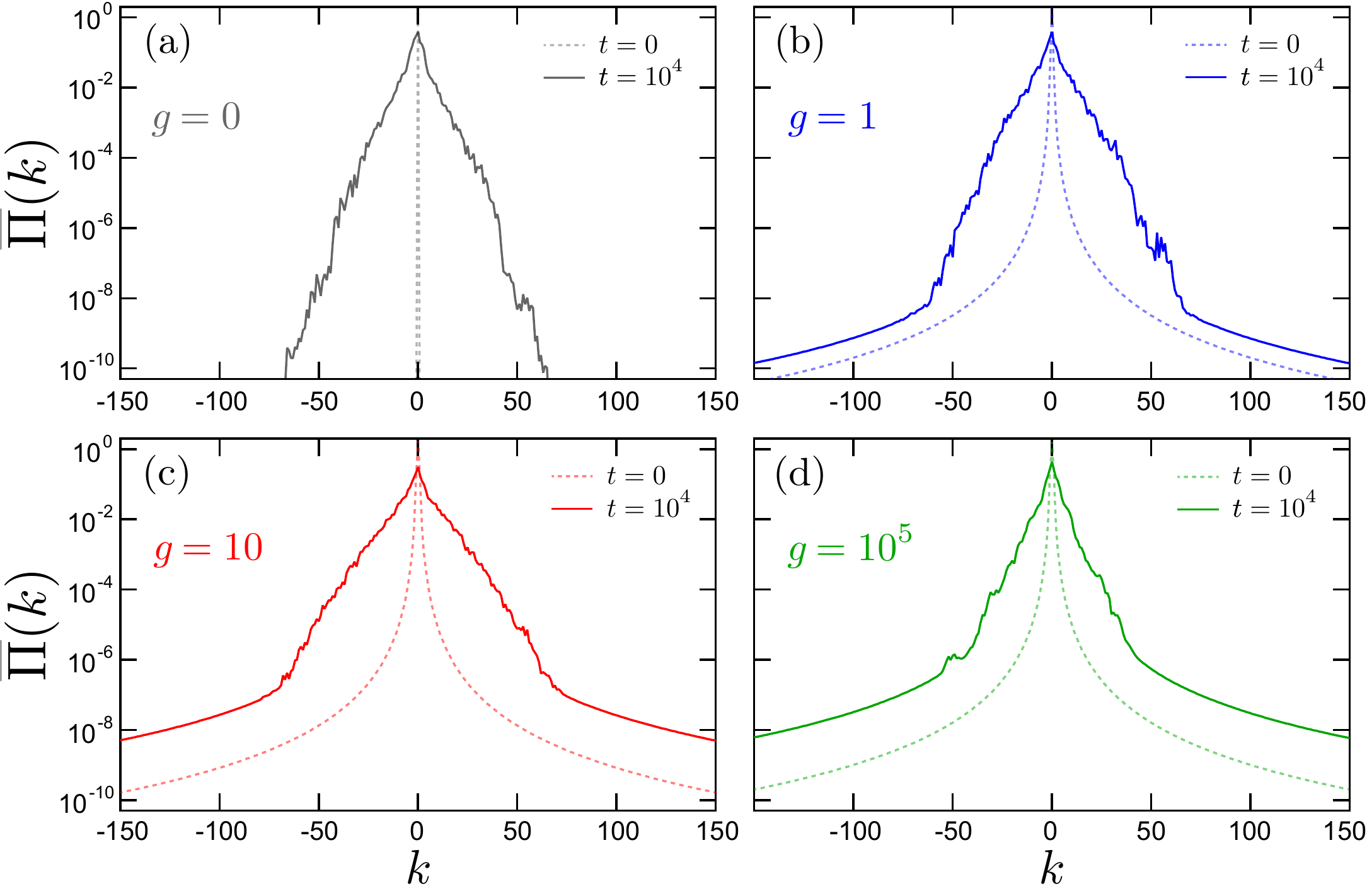}
	\caption{Averaged momentum distributions in the dynamically localized regime ($t=10^4$ kicks) for various values of the interaction parameter $g$. For $g\neq0$ all localized distributions (full lines) show the same common feature at large momenta (corresponding to power law $1/k^4$ tails), and an exponential decay at low momenta, with a $g$-dependent localization length. In the TG limit (d), we find the same typical localization length as the non-interacting case (a).}
	\label{fig4_Pi_k_c} 
\end{figure}

Leaving the details of the calculation to App. \ref{app_Pik}, the momentum distribution reads
\begin{equation}
\Pi_t(k) = \sum_{n,m,p} \big(c^n_m(t)\big)^*c^q_p(t)\Pi^{n,n}_{m,p}(k).
\end{equation}
with 
\begin{equation}
\Pi^{n,q}_{m,p}(k)=\delta_{n,q}\frac{A_m A_p  }{\pi  \left((2 k-n)^2-4
   k_m^2\right) \left((2 k-n)^2-4 k_p^2\right),
   }
\end{equation}
and
\begin{equation}
A_m=\frac{8 k_m \cos\left(\frac{\theta_m}{2}\right)}{\sqrt{\pi-\frac{\sin(\theta_m)}{2k_m}}}.
\end{equation}
At long time, the momentum distribution reads:
\begin{equation}
\overline\Pi(k) = \lim_{t\to\infty}\sum_{n,m,p} \overline{\big(c^n_m(t)\big)^*c^q_p(t)}\Pi^{n,n}_{m,p}(k).
\end{equation}
Since the coefficients $c^n_m(t)$ converge to a finite steady-state value, so does the momentum distribution, shown in Fig.~\ref{fig4_Pi_k_c}. The  distributions $\overline\Pi(k)$ display an exponential decay at small enough momenta, with a characteristic localization length which depends the interaction strength. However, at large momenta, the momentum distribution is dominated by a $k^{-4}$ tail, which is clearly visible for $g>0$.  This tail is a universal feature of interacting quantum systems, and already exists in the ground state (corresponding to the $t=0$ curves in Fig.~\ref{fig4_Pi_k_c} for $g>0$)\cite{Olshanii2003,Tan2008}. This behavior at large momenta is in sharp contrast with the non-interacting limit of the kicked rotor, where the momentum distribution decays exponentially.

The power law tail is charaterized by the so-called Tan contact, $\mathcal C= \lim_{k\to\infty} k^4\Pi(k)$.
Noting that $\Pi^{n,n}_{m,p}(k)$ decays at large momenta as $k^{-4}$ for all $n,m,p$, we find that:
\begin{equation}
\lim_{k\to\infty} k^4\overline\Pi(k)=\overline{\mathcal C},
\end{equation}
where
\begin{equation}
\overline{\mathcal C} = \lim_{t\to \infty}\sum_{n,m,p}\overline{\big( c^n_m(t)\big)^* c^n_p(t)}\frac{A_m A_p}{16\pi},
\end{equation}
is the effective Tan's contact in the dynamically localized regime. (We have checked that the contact obtained with the above equation describes very well the tail of the momentum distribution in Fig.~\ref{fig4_Pi_k_c}.)

\begin{figure}[t!]
	\centering
	\includegraphics[width=\columnwidth]{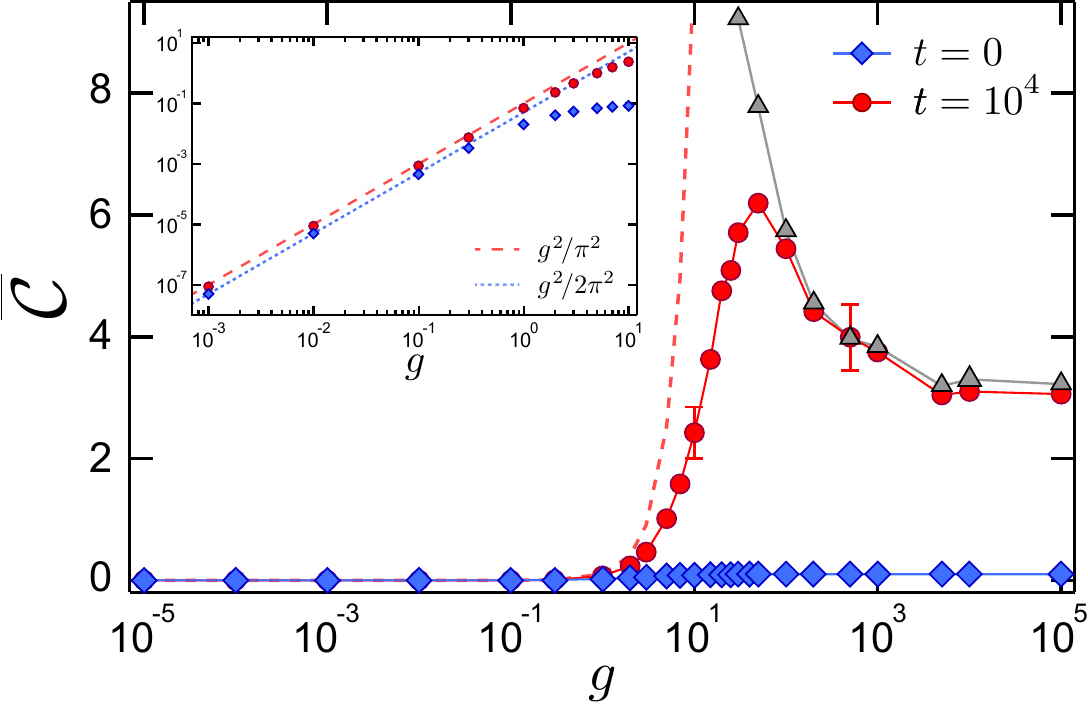}
	\caption{Evolution of the average Tan's contact $\overline{\mathcal{C}}$ as a function of the interaction parameter $g$, in the case of the initial Lieb-Liniger ground state (blue diamonds) and in the localized regime (red circles).  The dashed line corresponds to the analytical asymptotic behavior $\overline{\mathcal C}\simeq g^2/\pi^2$ in the weak interaction regime, and the triangles are obtained in the TG limit, using the analytical expression  $\overline{\mathcal C}\simeq \frac{2 \overline{E}_{\rm tot.}}{\pi^2\kb^2}$ (see text). Solid lines are guides to the eye. Statistical error bars are due to averaging over $\beta$ (typically 200 values). The inset shows the weak coupling regime in logarithmic scale. The lines are the analytical the ground state contact in this regime, at $t=0$ (dotted blue line: $\mathcal C\simeq g^2/2\pi^2$, see text), and at $t\gg t_{\rm loc.}$ (dashed red line).}
	\label{fig5_Tan_vs_c}
\end{figure}

This feature is also dependent of the value of interactions, and is captured in the evolution of Tan's contact shown in Fig.~\ref{fig5_Tan_vs_c}. At low interaction strengths, the value of the contact  in the localized regime $\overline{\mathcal{C}}$ (red) remains small, and is roughly proportional to its initial value (blue). Above a certain threshold of $g$, which is on the order unity, $\overline{\mathcal{C}}$ increases significantly. It reaches its maximum for $g\sim 50$, which is, maybe counter-intuitively, not where the energy is maximum (around $g=10$). It is worth to point out that, whereas for very large $g$ the energy decreases towards the same value as in the non-interacting case (see Fig.~\ref{Fig3-Etot_vs_c}), in the TG limit the Tan's contact saturates to a finite value. This difference is also clearly observed in Fig.~\ref{fig4_Pi_k_c}, bottom right panel. 

\begin{figure}[t!]
	\centering
	\includegraphics[width=\columnwidth]{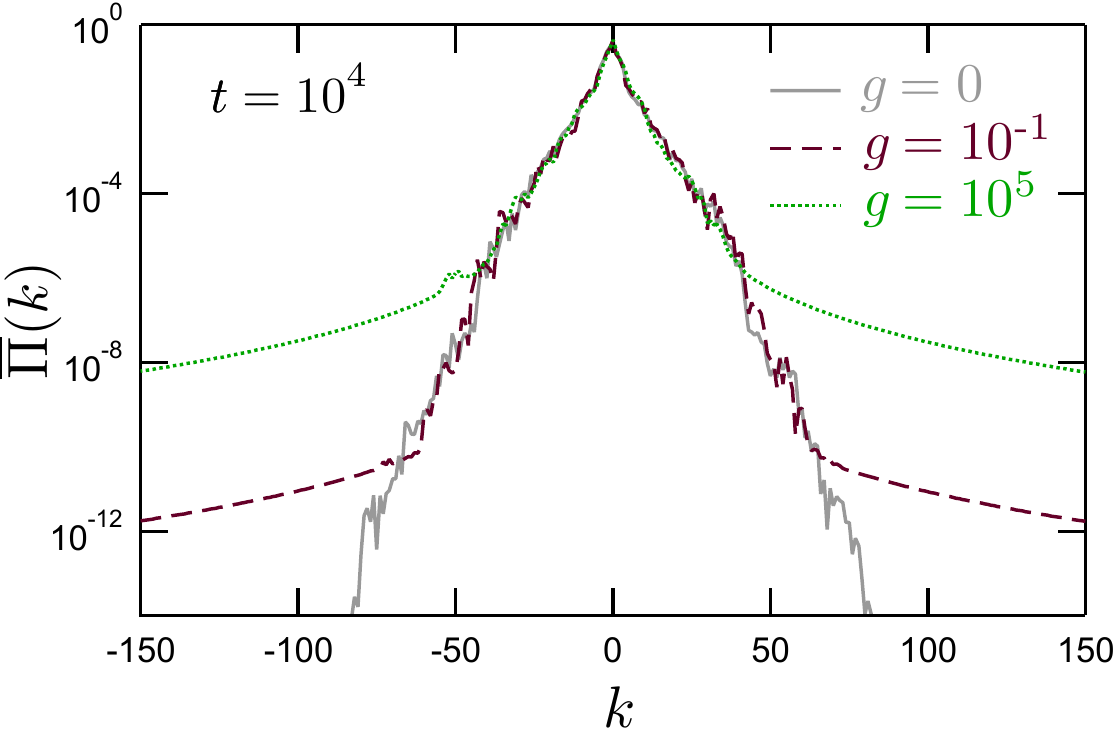}
	\caption{Averaged momentum distributions in the dynamically localized regime ($t=10^4$ kicks) for weak  ($g=10^{-1}$) and strong ($g=10^5$) interactions. The exponential decrease at small $k$ is almost identical, and very well described by the momentum distribution of non-interacting bosons ($g=0$). At larger momenta, the momentum distribution in the Tonks regime is dominated by Tan's contact. }
	\label{fig4_bis_Pi_k_c} 
\end{figure}

The shape of the momentum distribution can be understood quantitatively for weak ($g\to0$) and strong ($g\to\infty$) interactions. The details of the calculations are given in App.~\ref{app_asymptotic}, and we only use the results to discuss the momentum distribution and the contact in these two regimes.
Both in the weak and strong interactions limit, we find that the momentum distribution has typically two behaviors: i) at small enough momenta, it decays exponentially and is well approximated by the momentum distribution of two non-interacting bosons starting at zero-momentum; ii) at large enough momenta, the power law decay, $\overline\Pi(k)\simeq \overline{\mathcal{C}}/k^4$ dominates. These behaviors are shown in Fig.~\ref{fig4_bis_Pi_k_c}.
In the weak interaction limit, we find that $\overline{\mathcal C}\simeq g^2/\pi^2$ (compared to $g^2/2\pi^2$ in the ground state), whereas in the TG regime, we find $\overline{\mathcal C}\simeq \frac{2 \overline{E}_{\rm tot.}}{\pi^2\kb^2}$. These asymptotic formulas describe very well the contact in these two regimes, as can be seen in Fig.~\ref{fig5_Tan_vs_c}.

\section{Discussion \label{sec_discuss}}

Our results are in stark contrast with the conclusions of Qin et al. \cite{Qin2017}, who found for the same model and parameter range that interactions lead to delocalization. This affirmation was based on two results: i) By computing the variance of the momentum up to $5000$ kicks, they observed a somewhat increasing trend, which they interpreted as delocalization; ii) Their major argument was that the coefficient $c^n_m(t)$ behaved as $m^{-4}$ at long time (contrary to the exponential decay in the non-interacting limit), which they also interpreted as a sign of delocalization.

Concerning the first point, we note that their numerical simulation were not averaged, which makes it difficult to interpret the absence of localization (as can happen in the non-interacting QKR for some specific values of the parameters if not averaged over the quasi-momentum). Concerning the second point, we do agree with the $m^{-4}$ behavior of $c^n_m(t)$. However, this power law decay does not imply delocalization. Indeed, as we have shown above, the total energy (which has a term proportional to $\sum_{n,m} m^2  |c^n_m(t)|^2$) does saturate at long times. Furthermore, the coefficients converge to finite steady-state values. Finally, and more importantly, it is known that in some disordered model with power law (but short-range) hoping, corresponding here to $|U_{mp}^{nq}|\sim m^{-\mu}$ for large $m$ and fixed $n,q,p$, the states are localized as long as $\mu>3/2$ \cite{Moura2005}. Since the matrix element of the present problem decay with $\mu=4$, dynamical localization is therefore expected. To support this, we analyze in App.~\ref{app_toy_model} a modified QKR with matrix elements decaying as $m^{-4}$, and we show that indeed it dynamically localizes.

\section{Conclusions \label{sec_concl}}
We studied the outcome of dynamical localization with the kicked rotor model of two interacting bosons, and demonstrated its survival for arbitrary interaction strengths. The localization energy is found equal in the non-interacting (free bosons) and TG limits, and displays a non-monotonous behavior. Moreover, new features are predicted for the shape of the momentum distribution, namely the subsistence of an exponentially-localized `core', at low momenta, and the existence of a power law decay at large momenta -- a key characteristic of interacting quantum particles. Both features depend, yet in different manners, on the strength of the interaction.

An interesting question is the outcome of dynamical localization in the many-body limit. For interacting bosons in the TG limit, our localization argument still holds: the energy is rigorously equal to that of $N$ free fermions, and thus saturates at long times to a finite value, with the same localization time scale. This has already been predicted in~\cite{Rylands2020}. However, the nature of this localized state is still to be determined. While our work does not address the many-body momentum distribution, we expect our conclusions concerning the contact and power law tail at large momentum to be robust. This is especially relevant for future experimental observation of many-body dynamical localization. A comprehensive study of these aspects can be found in Ref.~\cite{Vuatelet2021}. Finally, it is an interesting question as whether the subdiffusion in momentum space, predicted by mean field methods, could be observed in a fully quantum kicked system, even on finite time window. This could indeed be the case in the weak interaction limit, which is known to be rather singular for the Lieb-Liniger model.

\section*{Acknowledgments}
We thank J.-C. Garreau for discussions, and acknowledge N. Krai for his involvement at an early stage of this work.
This work was supported by Agence Nationale de la Recherche through Research Grants MANYLOK No. ANR-18-CE30-0017 and QRITiC I-SITE ULNE/ ANR-16-IDEX-0004 ULNE, the Labex CEMPI Grant No.ANR-11-LABX-0007-01, the Programme Investissements d'Avenir ANR-11-IDEX-0002-02, reference ANR-10-LABX-0037-NEXT and the Ministry of Higher Education and Research, Hauts-de-France Council and European Regional
Development Fund (ERDF) through the Contrat de Projets \'Etat-Region (CPER Photonics for Society, P4S).

\bibliography{bibQKR_LL}

\appendix

\section{Calculation of the momentum distribution  \label{app_Pik}}

The momentum distribution is obtained from the OBRDM,
\begin{equation}
\Pi_t(k) = \frac{1}{2\pi}\int_0^{2\pi} dx\, \int_0^{2\pi} dy\, e^{i k (x-y)} \rho_t(x,y),
\end{equation}
with
\begin{equation}
\rho_t(x,y)=2\int_0^{2\pi}dz \,\Psi^*_t(x,z)\Psi_t(y,z).
\end{equation}

The OBRDM can be expressed as
\begin{equation}
\rho_t(x,y) = \sum_{n,m,q,p} \big(c^n_m(t)\big)^*c^q_p(t) \rho^{n,q}_{m,p}(x,y),
\end{equation}
with 
\begin{equation}
\rho^{n,q}_{m,p}(x,y) = 2\int_0^{2\pi}dz\, \big(\Phi^n_m(x,z)\big)^*\Phi_p^q(y,z).
\end{equation}
To get the momentum distribution, we need to compute $\Pi^{n,q}_{m,p}(k)$, the Fourier transform of $\rho^{n,q}_{m,p}(x,y)$. Noting that the invariance per translation of $\hat H_{LL}$ implies that $\Pi^{n,q}_{m,p}(k)$ vanishes if $n\neq q$, we obtain after a straight-forward though rather tedious calculation
\begin{equation}
\Pi^{n,q}_{m,p}(k)=\delta_{n,q}\frac{A_m A_p  }{\pi  \left((2 k-n)^2-4
   k_m^2\right) \left((2 k-n)^2-4 k_p^2\right)
   },
\end{equation}
where
\begin{equation}
A_m=\frac{8 k_m \cos\left(\frac{\theta_m}{2}\right)}{\sqrt{\pi-\frac{\sin(\theta_m)}{2k_m}}}.
\end{equation}

For a given state $|\Phi_m^n\rangle$, one can check that its momentum distribution $\Pi^n_m(k)$ obeys
\begin{equation}
\begin{split}
&\sum_k \Pi^n_m(k)=2,\\
&\sum_k k\kb \Pi^n_m(k)=n\kb,\\
&\sum_k \frac{\kb^2 k^2}{2}\Pi^n_m(k)=E_{\rm kin.}=E_m^n-E^n_{m,\rm int.},\\
\end{split}
\end{equation}
where $E^n_{m,\rm int.}=\langle \Phi_m^n|g\,\delta(\hat x_1-\hat x_2)|\Phi_m^n\rangle$ is the interaction energy.

From the above results, the momentum distribution reads
\begin{equation}
\Pi_t(k) = \sum_{n,m,p} \big(c^n_m(t)\big)^*c^q_p(t)\Pi^{n,n}_{m,p}(k).
\end{equation}

\section{Momentum distribution in the asymptotic regimes  \label{app_asymptotic}}

\subsection{Non-interacting limit}

In the limit $g\to 0$, the initial wave function is given by $\Phi_1^0(x_1,x_2)=(2\pi)^{-1}+\mathcal O(g)$, i.e. the two bosons start into the zero-momentum state. The dynamics is that of two independent bosons (up to $\mathcal O(g)$ corrections), and we can therefore assume that at long times, the two bosons are described by the same dynamically localized wave function of the non-interacting QKR $\psi_0(x)$, i.e.
\begin{equation}
\Psi(x_1,x_2) = \psi_0(x_1)\psi_0(x_2)+\mathcal O(g).
\end{equation}
It is then straightforaward to show that in the Lieb-Liniger basis, the coefficients $c_m^n$ are given in the localized regime by
\begin{equation}
\begin{split}
c_1^n &= \hat \psi_0\left(\frac n2\right)^2+\mathcal O(g),\\
c_{m>1}^n &= \sqrt{2} \hat\psi_0\left(\frac{n+m-1}2\right) \hat\psi_0\left(\frac{n-m+1}2\right)+\mathcal O(g),
\end{split}
\label{eq_c_psi}
\end{equation}
where $\hat f$ is the Fourier transform of the function $f$. 

In the weak interaction limit, we find that the coefficients $A_m$ that enter in the momentum distribution (see App.~\ref{app_Pik}) are such that 
\begin{equation}
\begin{split}
\frac{A_1}{\sqrt{\pi}  \left((2 k-n)^2-4
   k_1^2\right)} &= \sqrt{2}\delta_{n,2k}+\mathcal O(g),\\ 
\frac{A_{m>1}}{\sqrt{\pi}  \left((2 k-n)^2-4
   k_m^2\right)} &= \delta_{n,2k+m-1}+\delta_{n,2k-m+1}+\mathcal O(g),\\ 
\end{split}
\end{equation}
which immediately gives
\begin{equation}
\Pi(k)=2|\hat\psi_0(k)|^2+\mathcal O(g),
\end{equation}
as expected for free bosons.

However, for momenta very large compared to the localization length $p_{\rm loc}$ of the non-interacting QKR, $|\hat\psi_0(k)|^2$ is exponentially small compared to the $\mathcal O(g)$ corrections, and the momentum distribution is dominated by the contact,
\begin{equation}
\Pi(k)\simeq \frac{\mathcal C}{k^4}.
\end{equation}
In this regime, we find
\begin{equation}
\mathcal C=	\frac{g^2}{\pi^2}\sum_{n,m,p} a_m a_p (c_m^n)^* c_p^n+\mathcal O(g^3),
\end{equation}
with $a_1=1/\sqrt{2}$ and $a_{m>1}=1$, where we can use Eq.~\eqref{eq_c_psi} to the same accuracy. We can now use the fact that the phases of the QKR wave functions  are essentially random, such that when averaging over $\beta$, only the diagonal terms $p=m$ survive, i.e. $\overline{(c_m^n)^* c_p^n}\simeq \delta_{m,p}\overline{|c_m^n|^2}$.

We then obtain
\begin{equation}
\overline{\mathcal C}=\frac{g^2}{\pi^2}\left(1-\frac{1}{2}\sum_q \overline{|\hat\psi_0(q)|^4}\right)+\mathcal O(g^3).
\end{equation}
For localized state, we expect $\frac{1}{2}\sum_q \overline{|\hat\psi_0(q)|^4}\simeq \frac{1}{8 p_{\rm loc}}$ to be small and the contact is thus  
\begin{equation}
\overline{\mathcal C}\simeq \frac{g^2}{\pi^2}.
\end{equation}

In summary, the momentum distribution is decays exponentially as $2|\hat\psi_0(k)|^2$for $|k|\ll p_c$ and as a power law $g^2/(\pi^2 k^4)$ for $|k|\gg p_c$, where the cross-over scale is given by
\begin{equation}
2|\hat\psi_0(p_c)|^2 \simeq \frac{g^2}{\pi^2 p_c^4}.
\end{equation}

A similar calculation shows that the contact in the the ground state is $g^2/2\pi^2$.

\subsection{Tonks-Girardeau regime}

In the limit $g\to\infty$, thanks to the Bose-Fermi mapping, we can write the wave function of the bosons in the localized regime as 
\begin{equation}
\Psi(x_1,x_2)=\frac{{\rm sign}(x_1-x_2)}{\sqrt{2}}
\det\begin{pmatrix}
\psi_{+}(x_1) &\psi_{-}(x_1)\\ \psi_{+}(x_2)&\psi_{-}(x_2)
\end{pmatrix},
\end{equation}
where $\psi_{\pm}(x)$ are the wave functions of non-interacting fermions, evolving according the non-interacting QKR Hamiltonian, with anti-periodic boundary conditions. The initial condition is such that the two fermions start in the momentum state $p_\pm = \pm\frac 12$. At long time, $\hat\psi_\pm(q)$ are exponentially localized with  localization length $p_{\rm loc}$ similar to that of free bosons. In particular, for large enough $p_{\rm loc}$, we expect $|\hat\psi_\pm(q)|^2\simeq |\hat\psi_0(q)|^2$ where $\hat\psi_0(q)$ is the localized wave function of a boson starting at zero-momentum.

In the Lieb-Liniger basis, the coefficients $c_m^n$ are then given by
\begin{equation}
c_m^n=\sum_{\sigma=\pm1}\sigma  \hat\psi_{+}\left(\frac{n+\sigma m}{2}\right) \hat\psi_{-}\left(\frac{n-\sigma m}{2}\right).
\end{equation}
Therefore, the momentum distribution reads
\begin{equation}
\Pi(k)=\frac1{\pi^2}\tilde{ \sum_{p_1,p_2,q_1,q_2}}B_{p_1,p_2,q_1,q_2}(k)\hat\psi_+^*(p_1)\hat\psi_-^*(p_2)\hat\psi_+(q_1)\hat\psi_-(q_2),
\end{equation}
where
\begin{equation}
B_{p_1,p_2,q_1,q_2}(k)=\frac{(p_1-p_2)(q_1-q_2)}{(k-p_1)(k-p_2)(k-q_1)(k-q_2)},
\end{equation}
and the sum $\tilde{ \sum}_{p_1,p_2,q_1,q_2}$ is over half-integers such that $p_1+p_2=q_1+q_2$.

Upon averaging over $\beta$, we expect 
\begin{equation}
\overline{\hat\psi_+^*(p_1)\hat\psi_-^*(p_2)\hat\psi_+(q_1)\hat\psi_-(q_2)}\simeq \delta_{p_1,q_1} \delta_{p_2,q_2}\overline{|\hat \psi_+(q_1)|^2 }\,\overline{|\hat \psi_-(q_2)|^2 },
\end{equation}
 since the phases of two different localized state of the QKR are (almost) uncorrelated.

 The averaged momentum distribution reads
\begin{equation}
\overline \Pi(k)=\frac1{\pi^2}\sum_{q_1,q_2}\frac{(q_1-q_2)^2}{(k-q_1)^2(k-q_2)^2}\overline{|\hat \psi_+(q_1)|^2 }\,\overline{|\hat \psi_-(q_2)|^2 }.
\end{equation}

We have observed numerically that for small enough momenta, $\overline \Pi(k)$ is well described by
\begin{equation}
\begin{split}
\overline \Pi(k)&\simeq \overline{|\hat \psi_+(k)|^2 }+\overline{|\hat \psi_-(k)|^2},\\
&\simeq 2\overline{|\psi_0(k)|^2},
\end{split}
\end{equation}
where we have assumed that the width of the wave functions (given by $p_{\rm loc}$) is much larger than one to go from the first to the second line.
For large momenta we have $\overline \Pi(k)\simeq \overline{\mathcal C}/k^4$ with the averaged contact
\begin{equation}
\begin{split}
\overline{\mathcal C}&\simeq \frac1{\pi^2}\sum_{q_1,q_2}(q_1-q_2)^2\overline{|\hat \psi_+(q_1)|^2 }\,\overline{|\hat \psi_-(q_2)|^2 },\\
&\simeq \frac{2 \overline{E}_{\rm tot.}}{\pi^2\kb^2},
\end{split}
\end{equation}
where the averaged total energy is given by $\overline E_{\rm tot.}=\frac{\kb^2}{2}\sum_{q} q^2\left(\overline{|\hat \psi_+(q)|^2}+\overline{|\hat \psi_-(q)|^2}\right)$. To go from the first to the second line, we have assumed that the wave functions are broad enough such that we can neglect $\sum_q q \overline{|\hat \psi_\pm(q)|^2}$.

A cross-over scale between the exponential and power law decay of the momentum distribution can be defined similarly as in the weak interaction regime.

\begin{figure}[t!]
	\centering
	\includegraphics[width=0.7\columnwidth]{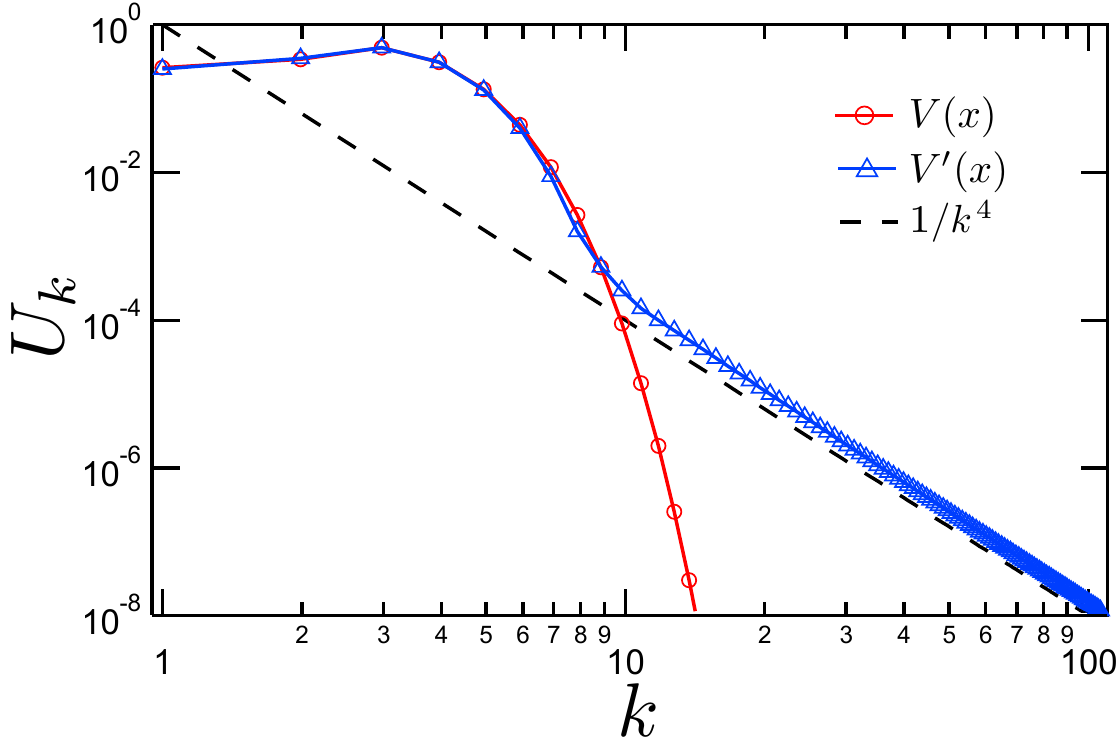}
	\caption{Coupling matrix elements of the standard (red circles) and modified (blue triangles) QKRs.}  
	\label{fig6_U_toy} 
\end{figure}

\section{Dynamical localization of a modified QKR \label{app_toy_model}}

We analyze a modified QKR model engineered such that the evolution operator decays as a power law similar to that of the kicked Lieb-Liniger gas, and we show that this power law behavior does not change the localization properties.

We introduce the toy model
\begin{equation}
\hat H' = \frac{\hat p^2}{2}+K V'(\hat x)\sum_n\delta(t-n),
\end{equation}
with $[\hat x,\hat p]=i\kb$, and the kick potential 
\begin{equation}
V'(x)= \frac{2x^4}{\pi^4}-\frac{4 x^2}{\pi^2}+1,
\end{equation}
for $x\in [-\pi,\pi[$, and $V'(x)$ is of period $2\pi$. This potential and its first and second derivative are continuous, whereas its third derivative is piece-wise continuous, which implies that its Fourier coefficients $\hat V_n$ decay as $n^{-4}$. The corresponding evolution operator over one period is
\begin{equation}
\hat U' = e^{-i\frac K\kb V'(\hat x)}e^{-i\frac{\hat p^2}{2\kb}},
\end{equation}
and by the same argument, one has
\begin{equation}
\lim_{|p'-p|\to \infty}\langle p'|\hat U'|\hat p\rangle \propto |p'-p|^{-4}.
\end{equation}
This behavior is demonstrated in Fig~\ref{fig6_U_toy}.

\begin{figure}[t!]
	\centering
	\includegraphics[width=\columnwidth]{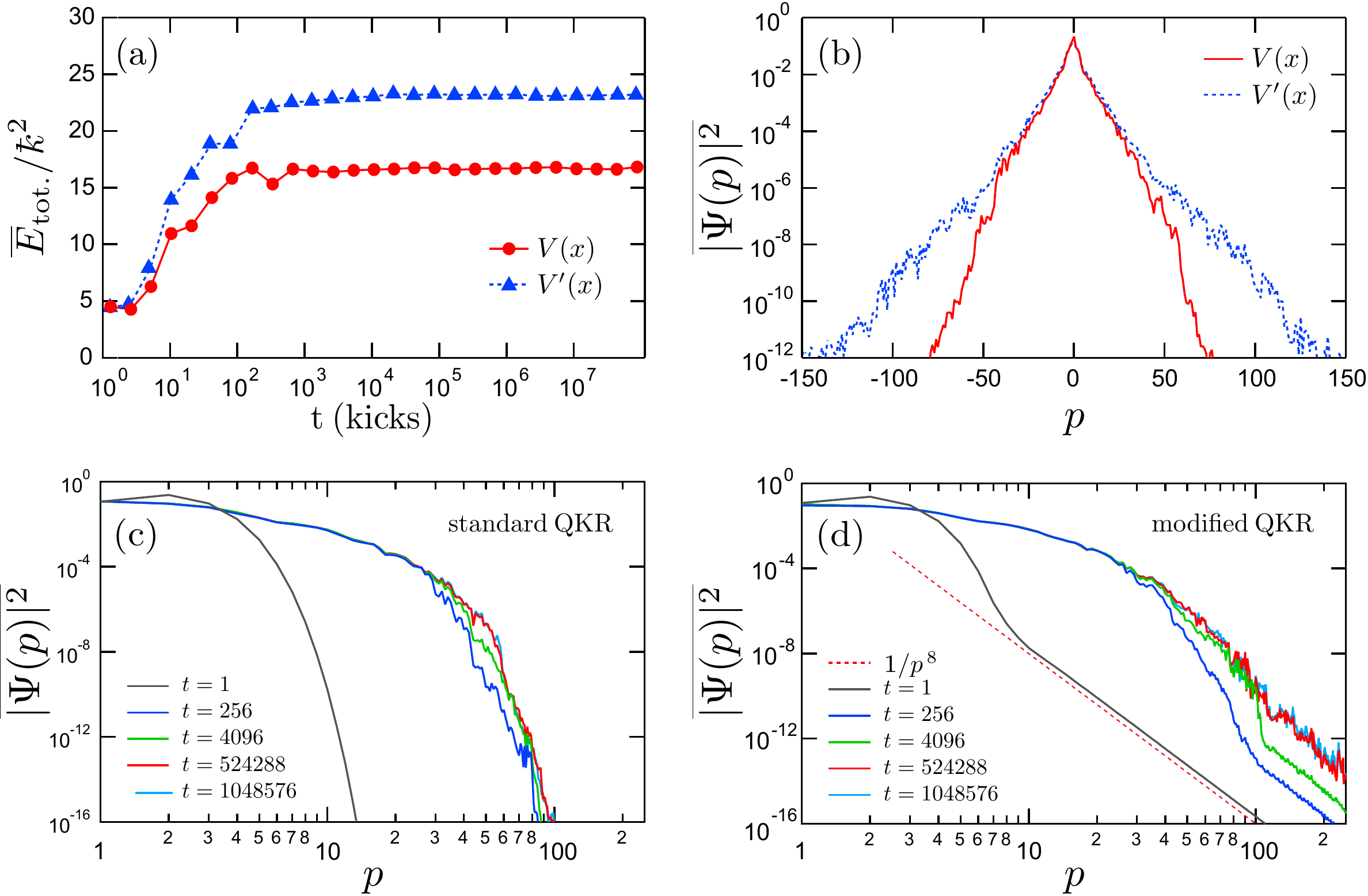}
	\caption{Comparison of the dynamics of the single-particle standard and modified QKRs ($K=3$, $\kb=1$): (a) saturation of the kinetic energy; (b) comparison of localized wave functions, shown for $t=2^{16}$ kicks; the (c) and (d) panels show, in log-log scales, wave functions at different times for the standard, respectively modified QKR ($t$ increases from left to right curves). For the modified QKR, the $1/p^4$ matrix elements lead to the population of a power law tail ($\propto 1/p^8$), present already after the first kick. However, the tail \textit{does} localize at long times, with a time scale much longer than that of exponentially-localized `core' of the momentum distribution (which dominates the localized kinetic energy).}
	\label{fig_Ec_toy} 
\end{figure}

The numerical analysis of this model is much simpler than that of the kicked Lieb-Liniger model, and one convinces oneself rather quickly that for generic values of the parameters (choosing $\kb$ not rational multiple of $\pi$ to avoid quantum resonances), the kinetic energy of the system always saturates at long times, see Fig.~\ref{fig_Ec_toy}. In the localized regime, we observe that similarly to the Lieb-Liniger case, the wave function take a steady-state shape, and decay as $|\langle p|\psi\rangle|^4 \propto p^{-8}$ in momentum space for large momenta, see Fig.~\ref{fig_Ec_toy}. However, this power law tail does not change the fact that the inverse partition ratio $P=\sum_p |\langle p|\psi\rangle|^4$ is always finite, which is a hallmark of localization. Because of the power law nature of the momentum coupling of $\hat{U}'$, the momentum distribution features a long power law tail even after a single kick. We note that the large-momentum power law tails localizes over longer time scales than the system energy, but still ends up localizing to a constant value.

\begin{figure}[t!]
	\centering
	\includegraphics[width=\columnwidth]{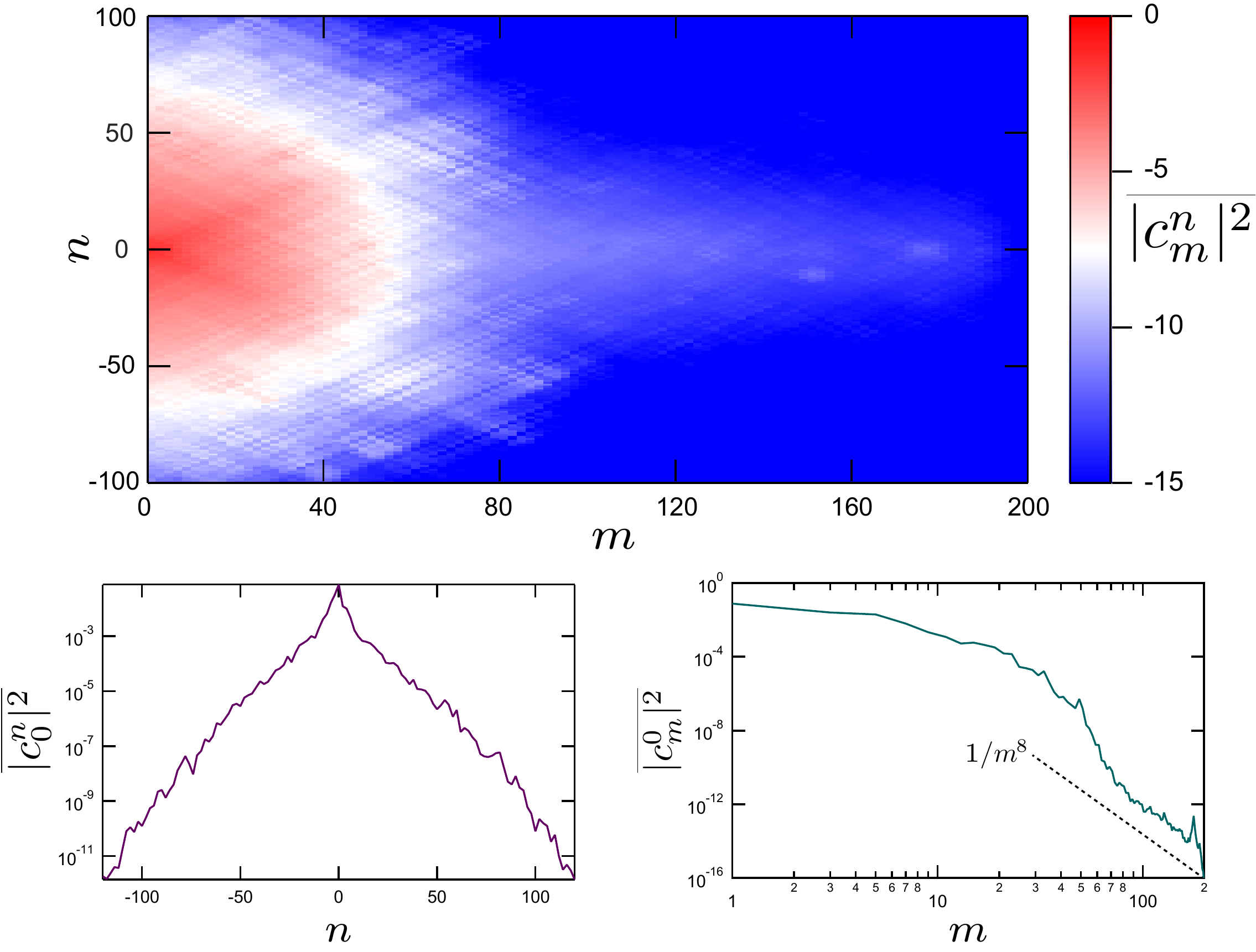}
	\caption{Localized two-body wave function in the Lieb-Liniger basis (top), at $t=2^{28}$ kicks ($K=3$, $\kb=1$, $g=1$). The wave function is exponentially localized in the center of mass direction $n$ (bottom left), and displays a long-range $1/m^8$ tail (bottom right), characteristic of the power law coupling (similar to the $V'(x)$ potential of the modified single-particle QKR).}
	\label{fig_LLwaveFcn} 
\end{figure}

To push the analysis further, we can also analyze the shape of the wave function in the Lieb-Liniger basis $|c_m^n|^2$. This is shown in Fig.~\ref{fig_LLwaveFcn}. While we observe an exponential localization in the center of mass direction $n$, the shape of the wave function coefficients $|c_m^n|^2$ display the characteristic power law $1/m^8$ tails along the relative momentum direction $m$.

\end{document}